\documentstyle[epsf,twocolumn,pre,aps,amssymb]{revtex}
\begin{document} 
\wideabs{    
\draft
\title{ Heat conduction and long-range spatial correlation in 1D 
models}
\author{ Xin Zhou and Mitsumasa Iwamoto}
\address{Department of Physical Electronics, Tokyo Institute of 
Technology, O-okayama 2-12-1, Meguro-ku, Tokyo 152-8552, Japan }
\date{\today}

\maketitle

\begin{abstract}
 Heat conduction in one-dimensional (1D) systems is studied based on an 
 analytical S-matrix method, which is developed in the mesoscopic electronic 
 transport theory and molecular dynamic (MD) simulations. It is found
 that heat conduction in these systems is related to spatial correlation of 
 particle motions. 
 Randomizations of scatterers is found to break the correlation, hence results 
 in normal thermal conduction. Our MD simulations are in agreement with 
 the theoretical expectations. The results are useful for an understanding 
 of the relation between heat conduction and dynamic instablities or other 
 random behavior in 1D systems.
\end{abstract}
}

\section{Introduction}
The study of heat conduction in one-dimensional (1D) systems is an 
interesting subject in the context of nonequilibrium statistical physics, 
which has been attracting much attention in recent years~\cite{Bonetto2000}. 
Most works aim at understanding the dynamic properties of heat 
transport 
in 1D systems~\cite{Casati1984,Lepri1997,Hu1998,Hatano1999,Alonso1999,Gendelman2000,Dhar2001,Li2002,Grassberger2002}. Although many modelling systems have 
been carefully studied and their thermal conductivities and temperature 
profiles calculated, the key dynamic properties of normal 
conduction is still unknown. Generally, normal heat conduction is 
specified by a diffuse-type motion, hence the investigation of random 
behavior of systems is necessary. It is well known that chaos can 
generate the needed random behavior in some systems. For example, thermal 
conductivity is characterized normal in the ding-a-dong 
model~\cite{Casati1984} and the Lorentz gas channels~\cite{Alonso1999}, due to 
their dynamic instability. 
However heat conduction is still considered abnormal in other systems, 
such as Fermi-Pasta-Ulam (FPU) model~\cite{Lepri1997} in spite of their 
dynamic instability. 
Recently, Li {\it et al.}~\cite{Li2002} studied three models with the same 
zero-Lyapunov exponents and found that heat conduction can be either 
normal or abnormal depending on details of the models. However they could not 
reach the conclusion that there is direct connection between dynamic
instability and normal heat conduction. From these studies, the most 
important and general conclusion is that the momentum conservation systems 
with non-zero presure have anomalous conductivity, whereas, anomalous 
conductivity does not always imply momentum conservation. 
Prosen and Campbell~\cite{Prosen2000} were the first to prove this conclusion. 
Recently, Narayan and Ramaswamy~\cite{Narayan2002} criticized that the 
original proof is questionable, but there is no strong objection to 
deny the conclusion. The general theoretical result will explain the 
abnormal heat transport found in FPU model, and may clarify the recent 
confused results obtained by numerical works in some models, such as 
diatomic Toda model. Heat conduction in the Toda model  
was thought as normal by Jackson {\it et al.}~\cite{Jackson1989} and 
by Garrido {\it et al.}~\cite{Garrido2001}, but was against 
the numerical results by Hatano~\cite{Hatano1999}, Dhar~\cite{Dhar2001}, 
Savin {\it et al.}~\cite{Savin2002} and Grassberger 
{\it et al.}~\cite{Grassberger2002}. This indicates that it is difficult to 
get definite conclusions about the macroscopic transport properties of 
nonlinear systems by merely using short-time numerical simulations in small 
systems. 

In this paper, based on both theoretical and numerical methods, we try to 
explain the required dynamic characteristics that guarantee normal 
heat conduction in 1D systems. By considering a 1D model with $N$ scatterers 
and noninteracting classic particles, we find that the classic model can be 
theoretically treated using S-matrix theory~\cite{Datta1995} 
developed in the mesoscopic electronic transport theory (METT). In METT, if 
the scattering is random, electronic transport will be incoherent, 
electronic phase correlation at the two ends of a 1D chain will be absent, 
and eventually the Ohm's law is observed. Similarly, in our model, defining 
a `phase' correlation, we find that normal heat conduction is 
characterized by the broken of phase correlation, where the randomization 
of the scatterers makes a main contribution. This conclusion is supported by 
some principle considerations and previous related researches and can be 
applied to more general systems. Our result indicates that a sufficient 
random characteristic of 1D systems is necessary to obtain normal heat 
conduction.
The random may be ensured by dynamic instablity, randomization of scatterers 
or other matters and could destroy long range correlations of particle 
motion along the 1D chains. Therefore, dynamic instability is not a necessary 
condition to guarantee normal transport if other random effects exist.  
 
\section{Theory}

\subsection{Model and Scattering Theory}

Consider a 1D classical chain with $N$ scatterers $s_{k}$, $(k = 1,
\cdot \cdot \cdot, N)$ and non-interactive particles, which are elastically 
transmitted or reflected at these scatterers, we place this chain between two 
heat baths. Without loss of generality, the length of chain ($L$) is set as 
$N$. If the transmission coefficient ($t_{k}$) of each scatterer is a number
between $0$ and $1$, it is a typical random walk process, hence heat 
conduction is normal and independent of the kind of heat baths. However, here 
we suppose that scatterers periodicaly turn on or off in time, so particles 
will completely transmit through a scatterer in some time ranges, but reflect 
at the scatterer in the other time ranges. So the transmission coefficients 
are function of time $\phi$, ($t_{k}(\phi)$). It is a typically deterministic 
system. Here, we set the period $t_{k}(\phi)$ to $1$ and the average 
transmit coefficient to $1/2$. We suppose that there are different initial 
time shifts ${\delta}_{k}$ for different scatterers, then we have 
$t_{k}(\phi) = t(\phi - {\delta}_{k})$, and 
\begin{eqnarray}
t(\phi) = \left\{ 
\begin{array}{@{\,}ll}
1, & \mbox{$ 0 \le \phi < 1/2 $}\\
0, & \mbox{$ 1/2 \le \phi < 1 $}.
\end{array}
\right.
\end{eqnarray}
The term $\phi$ as well as ${\delta}_{k}$ are thought as `phases'. Another 
parameter of the $k$th scatterer is its positions $x_{k}$. 
$\{{\delta}_{k}\}$ and $\{x_{k}\}$ will determine the properties of the model.
Actually, the model is very similar to the Lorentz channel 
model~\cite{Alonso1999} 
or the Ehrenfest gas channel model~\cite{Li2002}. 

We define the average of the particle current density $J(x,v,t)$ on the
model,
\begin{eqnarray}
     j(x,v,\phi;n) = {1 \over M} \sum_{m=0}^{M-1} J(x,v,\phi + m + n),
\end{eqnarray}
where $n$, $m$ are integer numbers, $M$ is a large integer number, and 
$0 \le \phi<1$. In steady state, $j(x,v,\phi;n)$ is independent of $n$, 
and is noted as $j(x,v,\phi)$. Due to the particle current conservation, 
if $x$ is between two nearest neighbor scatterers, $x_{k}$ and $x_{k+1}$, 
$J(x,v,t)$ is written as $J_{v}(t-x/v)$. Hence we have,
\begin{eqnarray}
     j(x,v,\phi) = \sum_{m} \ j^{(k)}_{m}(v) 
     &\exp& \{2 m \pi i \ (x/v - \phi)\},  \\
     &if& \ \  x_{k} < x < x_{k+1}.  \nonumber
\end{eqnarray}
The heat current density $J_{u}(x)$ and temperature profile ${\cal T}(x)$ 
can be writen as,
\begin{eqnarray}
J_{u}(x) = {1 \over 2} \int_{0}^{\infty} d v v^{2} 
[j^{(k)}_{0}(v) - j^{(k)}_{0}(-v)], 
\end{eqnarray}
and 
\begin{eqnarray}
{\cal T}(x) = \frac{\int_{0}^{\infty} d v v 
[j^{(k)}_{0}(v) + j^{(k)}_{0}(-v)]}{\int_{0}^{\infty} d v v^{-1}
[j^{(k)}_{0}(v) + j^{(k)}_{0}(-v)] },
\end{eqnarray}
respectively, while $ x_{k} < x < x_{k+1}$. 
 
Based on the properties of scatterers, we easily obtain the following 
scattering formula,
\begin{eqnarray}
\left(
\begin{array}
{@{\,}c@{\,}}
{\hat j}^{(k)}(v) \\
{\hat j}^{(k-1)}(-v)\\
\end{array}
\right)  =  
\left(
\begin{array}
{@{\,}cc@{\,}}
{\hat t}^{(k)}(v) & {\hat r}^{\prime (k)}(v) \\
{\hat r}^{(k)}(v) & {\hat t}^{\prime (k)}(v) \\
\end{array}
\right)  \ 
\left(
\begin{array}
{@{\,}c@{\,}}
{\hat j}^{(k-1)}(v) \\
{\hat j}^{(k)}(-v)\\
\end{array}
\right)  
\label{scatter1}
\end{eqnarray} 
where ${\hat t}$, ${\hat r}$, ${\hat t}^{\prime (k)}$, and 
${\hat r}^{\prime (k)}$ are nothing but S-matrix elements. 
Since this is a multi-mode scattering problem, each S-matrix element is an
infinite dimensional matrix. For example, the element
$t^{(k)}_{mn}$ of \ ${\hat t}^{(k)}$ means a coefficient which the mode 
$m$ at the left transmit the scatterer $k$ to mode $n$ at the right, 
\begin{eqnarray}
 t^{(k)}_{mn}(v) = t^{(m-n)} 
 \exp \{ -2 \pi i (m-n) (x_{k}/v - {\delta}_{k}) \},
\end{eqnarray}
where $t^{(p)} = \int_{0}^{1} d \phi \  t(\phi) e^{2 p \pi i \phi}$, 
is the $p$th Fourier's expanded coefficient of transmission function 
$t(\phi)$. Similarly, other matrixes can be writen as, 
\begin{eqnarray}
r^{(k)}_{mn}(v) &=& r^{(m-n)}_{k} \exp [-2 \pi i (m+n) x_{k}/v] , \nonumber \\
t^{\prime (k)}_{mn}(v) &=& t^{(m-n)} 
\exp [-2 \pi i (m-n) (x_{k}/(-v) - {\delta}_{k})],  \\
r^{\prime (k)}_{mn}(v) &=& r^{(m-n)}_{k} \exp [-2 \pi i (m+n) x_{k}/(-v)], 
\nonumber
\end{eqnarray}
where $r^{(p)}_{k} = ({\delta}_{p,0} - t^{(p)})$ 
$\exp (2 p \pi i \ {\delta}_{k} )$. 
  
For whole $N$ scatterers, we have, 
\begin{eqnarray}
\left(
\begin{array}
{@{\,}c@{\,}}
{\hat j}^{(N)}(v) \\
{\hat j}^{(0)}(-v)\\
\end{array}
\right) \ \  = \ \ 
\left(
\begin{array}
{@{\,}cc@{\,}}
{\hat T}(v) & {\hat R}^{\prime}(v) \\
{\hat R}(v) & {\hat T}^{\prime}(v) \\
\end{array}
\right)  \ 
\left(
\begin{array}
{@{\,}c@{\,}}
{\hat j}^{(0)}(v) \\
{\hat j}^{(N)}(-v)\\
\end{array}
\right) 
\label{Nscattering}
\end{eqnarray}
where ${\hat T}$, ${\hat R}$, ${\hat T}^{\prime}$ and ${\hat R}^{\prime}$ 
are whole transmitting and reflecting matrix of all scatterers from left to 
right and from right to left, respectively. ${\hat j}^{(0)}$ and 
${\hat j}^{(N)}$ correspond to the current densities at the left
and right ends, respectively.
 
Since the parameters of the system ($x_{k}$ and $\delta_{k}$ ) determine the 
S-matrix, we study heat conduction of models with different parameters. 
Without loss of generality, we set 
\begin{eqnarray}
\delta_{k} &=& c * R, \label{randomphase} \\
x_{k} &=& k - 0.5 + d * (R - 0.5), 
\label{randomposition}
\end{eqnarray}
where $R$ is a random number uniformly distributed between $0$ and $1$, 
$c$ and $d$ are the magnitude of disorder in scattering phases and positions, 
respectively. If both $c$ and $d$ equal to zero, it is a periodic scattering 
system, otherwise, it is a disordered system.
For the latter, we can consider two kinds of different disordered systems: (1)
the dynamic random system (DRS), where $x_{k}$ or $\delta_{k}$ are 
random in time; (2) the static random system(SRS), where random
$x_{k}$ or $\delta_{k}$ are fixed rather than depending of time in each 
realizations of system. We are interested in the average properties of many 
realizations. DRS and SRS correspond to the phonon scattering and impure 
scattering in electronic transport, respectively. Obviously, the disorder of 
$x_{k}$ or $\delta_{k}$ will induce some random phases in the S-matrix, it 
may affect the transport property of the system.

The combining S-matrices of $N$ scatterers can be easily written according to 
the well known rules generated in METT: it is written as a summation of many 
terms, where each term is depicted by a `Feynman path'~\cite{Datta1995}.
For example, the combining transmitting matrix of any two parts of scattering
$s^{(1)}$ and $s^{(2)}$ is,
\begin{eqnarray}
{\hat t}(12) &=& {\hat t}^{(2)} [ I - {\hat r}^{{\prime} (1)} 
{\hat r}^{(2)} ]^{-1} {\hat t}^{(1)}, \nonumber \\ 
&=& {\hat t}^{(2)} [ I + {\hat r}^{{\prime} (1)} {\hat r}^{(2)} + 
\cdot \cdot \cdot ] {\hat t}^{(1)}.
\label{combinscatter}
\end{eqnarray}
Generally, we have $T_{mn}(N) = \sum_{P} A_{P}$, where each $A_{P}$ is a 
complex number contributed from path $P$ which starts from mode $n$ at 
the left end, to mode $m$ at the right end. If $t_{k}(\phi)$ of each 
scatterer is independent of time $\phi$, then there is only one mode, 
$T = \sum_{P_{0}} A_{P_{0}}$, where $P_{0}$ are all the `Feynman' paths with 
mode $0$ for all scatterers. We easily obtain,
\begin{eqnarray}
T(N) = ( \sum \frac{1 - t_{i}}{t_{i}} + 1 )^{-1}
\sim \frac{t}{N (1-t) + t}.
\end{eqnarray}
While $N \rightarrow \infty$, we have $T(N) \sim 1/N$, satisfying the 
Fourier's law, which is the expectation of the random walk process. 

\subsection{Heat Baths}
  
Before going on to the study of the transport properties of the model, we 
analyze the effects of heat baths. Here, the temperatures of the left and 
right heat baths are noted as $T_{1}$ and $T_{2}$, respectively. Many kinds
of heat baths can be selected. For example, we can choose complete-reflecting 
heat baths: as a particle hits a heat bath, it will be reflected with a 
velocity distribution $P_{T}(v)$, where $T$ is the temperature of the heat 
bath. $P_{T}(v)$ can be of the Maxwellian distribution, 
$P_{T}(v) = v/T \ {\exp}(-v^{2}/2T)$, or of a single velocity distribution, 
$P_{T}(v) = {\delta} ( v - \sqrt{T})$, or other forms. 
In this case, we have completely reflecting conditions at the boundaries:
\begin{eqnarray}
j^{l}_{m}(v) &=& J^{l}_{m}  \ P_{T_{1}}(v) \\
j^{r}_{m}(-v) &=& J^{r}_{m}  \ P_{T_{2}}(v),
\end{eqnarray}
where $v$ is larger than $0$, $j^{l}_{m}(v)$ / $j^{r}_{m}(-v)$  
corresponds to the currents entered into the 1D system from the left/right
heat bath ($x=0/L$), which is a different phase from 
$j^{0}_{m}(v)$ / $j^{N}_{m}(-v)$ which are used in eq. (\ref{Nscattering}).
$J^{l}_{m} = \int_{0}^{\infty} d v j^{l}_{m}(-v)$ and 
$J^{r}_{m} = \int_{0}^{\infty} d v j^{r}_{m}(v)$.
Actually, $J^{l}_{m}$ and $J^{r}_{m}$ are the Fourier expanded coefficients 
of the current densities $J^{l}(\phi)$ (at the left end, $x=0$) and 
$J^{r}(\phi)$ (at the right end, $x=L$) of 1D chain, respectively. 
From eq. (\ref{Nscattering}), we easily obtain,
\begin{eqnarray}
\left(
\begin{array}
{@{\,}c@{\,}}
{\hat j}^{r}_{m}(v) \\
{\hat j}^{l}_{m}(-v) \\
\end{array}
\right)  =  
\left(
\begin{array}
{@{\,}cc@{\,}}
\tilde{T}_{mn}(v) & \tilde{R}^{\prime}_{mn}(v) \\
R_{mn}(v) & \tilde{T}^{\prime}_{mn}(v) \\
\end{array}
\right)  \ 
\left(
\begin{array}
{@{\,}c@{\,}}
{\hat j}^{l}_{n}(v) \\
{\hat j}^{r}_{n}(-v) \\
\end{array}
\right)  
\end{eqnarray}
where there are some phase differences between the matrix elements 
and the original S-matrix elements in eq. (\ref{Nscattering}), 
\begin{eqnarray}
\tilde{T}_{mn}(v) &=& \exp (2 m \pi i L/v ) \ T_{mn}(v) \nonumber \\
\tilde{T}^{\prime}_{mn}(v) &=& T^{\prime}_{mn} \  \exp (2 n \pi i L/v ) \\
\tilde{R}^{\prime}_{mn}(v) &=& \exp (2 m \pi i L/v ) \  R^{\prime}_{mn} \ 
\exp (2 n \pi i L/v ) \nonumber
\end{eqnarray}
Then, we know $J^{l}_{m}$ and $J^{r}_{m}$ are not arbitrary, 
but satisfying the following equation,
\begin{eqnarray}
\left(
\begin{array}
{@{\,}cc@{\,}}
<R_{mn}>_{1} - {\delta}_{mn} & <\tilde{T}^{\prime}_{mn}>_{2} \\
<\tilde{T}_{mn}>_{1} & <\tilde{R}^{\prime}_{mn}>_{2} - {\delta}_{mn} \\
\end{array}
\right)  \ 
\left(
\begin{array}
{@{\,}c@{\,}}
J^{l}_{n} \\
J^{r}_{n} \\
\end{array}
\right)  = \ 0
\label{bscatter}
\end{eqnarray}
where $<f_{mn}>_{i}$, ($i=1$, or $2$ ), means the average 
value of 
$f_{mn}(v)$ under the velocity distribution function $P_{T_{i}}(v)$.
The physical meaning of eq.(\ref{bscatter}) will be easily understood. 
In fact, we can treat the right boundary as $(N+1)$th scatterer 
( complete reflecting), $J^{l}_{m}$ must satisfy the linear equation,
 $(R^{Nr}-I) J^{l} = 0$, 
 where $R^{Nr}$ is the total reflecting matrix of all $N$ scatterers 
 as well as the right heat bath. From Eq. (\ref{bscatter}), we have 
\begin{eqnarray}
R^{Nr} = <R>_{1} + <\tilde{T}^{\prime}>_{2} 
[I - <\tilde{R}^{\prime}>_{2}]^{-1} <\tilde{T}>_{1}.
\end{eqnarray}
Since 
\begin{eqnarray}
[I - R]^{-1} = I + R + R R + \cdot \cdot \cdot,
\end{eqnarray}
we find that the result is nothing but the summation of `Feynman path'.
Similarly, we also have the formula of $R^{lN}$ and equation 
$(R^{lN} - I) J^{r} = 0$. If there is only one linear independent solution 
of the equation, we will get a unique steady state, (a free constant 
$J^{l}_{0}$ is decided by the particle density and average 
temperature of system). 
 
However, the asymptotic $N$ dependence of the heat current is only determined 
by the $N$ dependence of transmission coefficient $|{\hat T}|$, which is 
independent of the details of heat baths. 
In this paper, we only consider a simpler case, replacing with 
the completely reflecting heat baths: particles uniformly enters into the 
system from two heat baths at time ($\phi$) with the single velocity 
distribution ${\delta} ( v - \sqrt{T})$, therefore the heat current is 
simply proportional to the transmission coefficient $T_{00}$.
When calculating ${\cal T}(x)$, we choose another uniform 
incoming current from the right heat bath
at the same time, making the total particle current zero. 

\subsection{Phase Correlation}

It is well established that, if the scattering is random, the contributions 
from 
different Feynman paths are incoherent in electronic transport, therefore, 
we can find ohm's law and normal electric conductivity. But if there are 
some long range correlations between scatterers, the electronic 
phase-relaxation length may be larger than the length of the system, the 
transport will be coherent and we can find anomalous electric conductivity. 
So the existence of the phase correlation between electrons at the two ends 
can be used to judge whether the transport be coherent or not, hence whether 
the conductivity be abnormal or normal. Comparing the results, we 
expect a similar relationship between heat conduction and a long range 
correlation of particles in 1D systems. We also expect the random 
charactaristic of scatterers to be responsible for the correlations. 

For any incoming particle current from the left heat bath, $J_{+}^{l}(\phi)$, 
we have the transmit current density at the right 
boundary $J_{+}^{r}(\psi) = \int T(\psi, \phi) J_{+}^{l}(\phi) d \phi$, where 
$T(\psi, \phi)$ is the transmit function, 
\begin{eqnarray}
T(\psi,\phi) = \sum_{mn} \exp (- 2 m \pi i \psi) \ T_{mn} \ 
\exp (2 n \pi i \phi)
\end{eqnarray}
Similarly, the reflecting function is $R(\psi,\phi)$. 
Since for any $\phi$, $\int [ T(\psi,\phi) + R(\psi,\phi) ] d\psi = 1$,
so $R_{0n} = {\delta}_{0n} - T_{0n}$. 
 
The probability that we observe a current with phase $\psi$ at the 
right end and a current with $\phi$ at the left end is
$W_{2}(\psi,L;\phi,0) = T(\psi,\phi) J_{+}^{l}(\phi)$.
We define a normalized two-point current distribution function as
$f_{2}(\psi,\phi) = W_{2}(\psi,N;\phi,0)/A$, 
where $A$ is a normalized constant, 
\begin{eqnarray}
A = \int W_{2}(\psi,L;\phi,0) d \phi \ d \psi = \int J_{+}^{r}(\psi) d \psi.
\end{eqnarray}
The left and right normalized distributions are
\begin{eqnarray}
f_{l}(\phi) &=& \int f_{2}(\psi, \phi) d\psi, 
\end{eqnarray}
and
\begin{eqnarray}
f_{r}(\psi) &=& \int f_{2}(\psi,\phi) d \phi, 
\end{eqnarray}
respectively.
  
Based on these distributions, we define a motion correlation of 
particles at two ends of the 1D chain as,
\begin{eqnarray}
D =\frac{\langle \phi \psi {\rangle}_{2} - \langle \phi {\rangle}_{l}
 \langle \psi {\rangle}_{r}}{\sigma_{l} \sigma_{r}},
\end{eqnarray}
where $\langle \cdot \cdot \cdot {\rangle}_{i}$ ($i=2$, $l$ or $r$) means the 
average value under the distribution functions $f_{2}(\psi,\phi)$, 
$f_{r}(\psi)$ and $f_{l}(\phi)$, respectively.
$\sigma_{l/r} = \sqrt{(\langle \phi^{2} {\rangle}_{l/r} - \langle \phi 
{\rangle}_{l/r}^{2})}$ is the 
width of the current distribution at the left/right end. Hence, $D = 0$ 
implies the broken of the correlation and $T(\psi,\phi) = F(\psi) G(\phi)$ 
(or $T_{mn} = F_{m} G_{n}$). Here, $D$ corresponds to the electronic phase 
correlation at both ends of 1D electronic transport system, and 
the particle velocity play the role of wavevector of electron. 

\subsection{Theoretical Results}

Based on the rules of the `Feynman' paths on scattering theory, the 
transmission matrix $T_{mn}$ can be writen as, 
\begin{eqnarray}
T_{mn} = \sum_{pq} t^{(N)}_{mp} t^{(1)}_{qn} B_{pq}(N),
\end{eqnarray}
where $t^{(N)}$ and $t^{(1)}$ are the transmission matrix of the scatterer 
${\mathbf N}$ and the scatterer ${\mathbf 1}$, respectively. 
$B_{pq}(N)$ is the sum of all `Feynman' paths which start
from mode $q$ of the scatterer ${\mathbf 1}$, end to mode $p$ of the scatterer 
${\mathbf N}$. Expanding $B_{pq}(N)$ in large $N$, the lead term is noted as
$\alpha_{pq}/N^{\gamma}$, therefore we have the first result that the 
asymptotic $N$ dependence of $T_{mn}$ is independent on $m$ and $n$, 
\begin{eqnarray}
T_{mn} = T(N) h_{mn},
\label{transmitmn}
\end{eqnarray} 
where $T(N) \sim 1/N^{\gamma}$, is a simple notation of $T_{00}$ and
$h_{mn}$ is independent of $N$. 
If the correlation is absent, $T_{mn} = T(N) f_{m} g_{n}$.

For DRS, due to the random time-dependent positions/phase of the 
scatterers, the contribution from different Feynman paths are not coherent
with each other, so they can be first averaged in time, then be summed up. 
This indicates that only $0$-mode contribute to the total transmission 
coefficient, so 
the heat conduction shall be normal. We also easily know that the correlation 
is absent in DRS due to the incoherent Feynman contribution. 
 
\section{Numerical Simulation}

\begin{figure}
\centerline{\epsfxsize=10cm \epsfbox{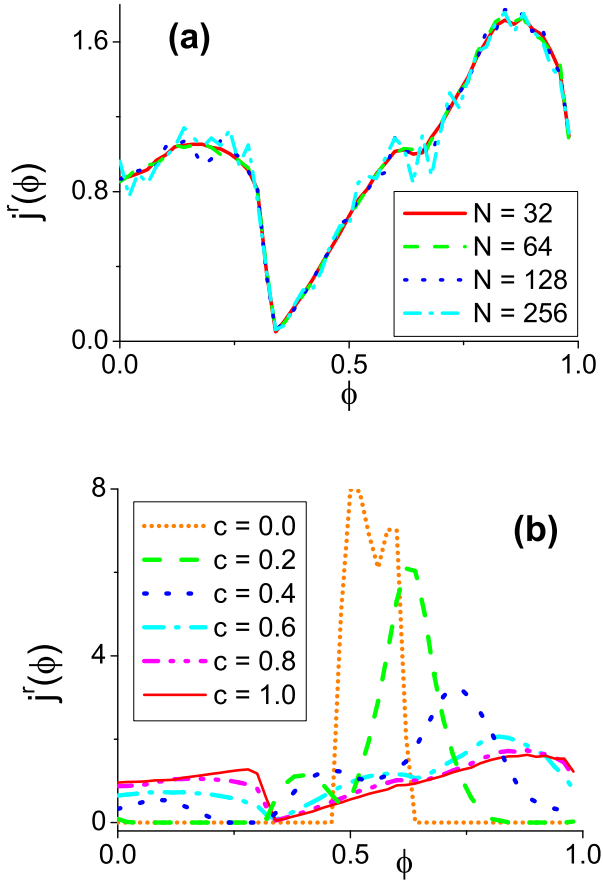}}
\caption{ The particle current distribution $J^{r}(\phi)$ at the right end 
in DRS with random phases: {\bf (a)} $J^{r}(\phi)$ of systems with different 
$N$ are identified each other, where $c=0.8$. {\bf (b)} 
the distribution
$J^{r}(\phi)$ of DRS with different $c$. $c=0$ means a periodic system. 
\label{fig1}}
\end{figure}

\begin{figure}
\centerline{\epsfxsize=10cm \epsfbox{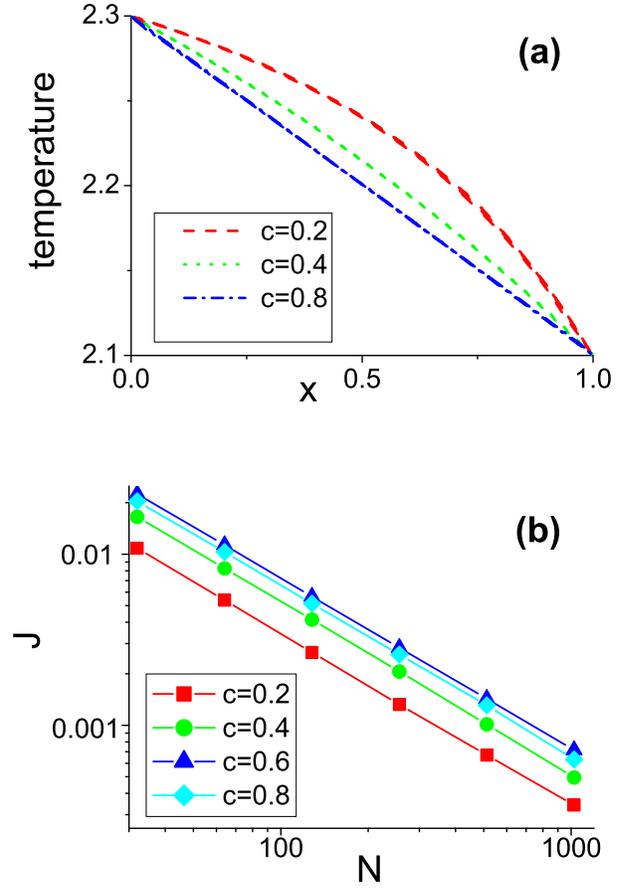}}
\caption{ The results of DRSs with random phase: {\bf (a)} Temperature 
profile, where $x=X/N$. For every 
$c$, we show four temperature profiles, the corresponding chain length are 
$32, 64, 128$ and $256$, respectively. They are almost identified each other. 
{\bf (b)} Heat current $J$ versus $N$.
\label{fig2}}
\end{figure}

In this section, by using a uniform inputing current with single velocity 
$v_{1} = \sqrt{T_{1}}$ from the left end, we numerically simulate
the transmit coefficient $T(N)$, the distribution function $f_{r}(\psi)$, 
$f_{l}(\phi)$ and the temperature profile ${\cal T}(x)$ in different systems.
First, we find that both $f_{r}(\psi)$ and $f_{l}(\psi)$ are independent of 
$N$ for all models including the periodic scattering system, DRS and SRS with 
the same disordered magnitude ($c$ and $d$). The results verify very well 
our theoretical expectations in eq. (\ref{transmitmn}) and indicate 
that $h_{mn}$ is only dependent on the disordered magnitude. As $c$ (or $d$) 
increases, $f_{r}(\psi)$ and $f_{l}(\phi)$ are flatter, but they are not 
uniform even though the disordered magnitudes arrive at their maximum 
value ($1$). For example, the results in 
 DRS are shown in Fig. \ref{fig1}, where $J^{r}(\psi)$ is the 
unnormalized current distribution function at the right end. 
Then, we show the temperature profiles ${\cal T}(X/N)$ of DRS 
in Fig. \ref{fig2}(a), which are found to being independent on $N$ but 
dependent on the disorder magnitude $c$ (or $d$). It indicates 
that $d{\cal T}/dX$ is proportional to $1/N$. The heat current
$J_{u}$ $\sim 1/N^{\gamma}$ is shown in Fig. \ref{fig2}(b), and fitted 
$\gamma \approx 0.9989 \pm 0.0045$. Therefore the heat conduction is normal 
in DRS with disordered scattering phases, which verifies our theoretical 
expectation. In this case, our obtained spatial correlation $D$ is very 
small but with the same order statistic error. For a better correlation 
estimation, we used the non-uniform input currents $J_{+}^{l}(\phi)$ 
with different distributions and found the right-end distribution functions 
$f^{r}(\psi)$ independent of the selection of $J_{+}^{l}(\phi)$
(not shown). This indicates that the transmission function $T(\psi,\phi)$ can 
be writen as $F(\psi) G(\phi)$, hence $D=0$. For DRS with disordered 
scattering positions, the obtained results are similar as 
that of DRS with disordered phases. 

\begin{figure}
\centerline{\epsfxsize=10cm \epsfbox{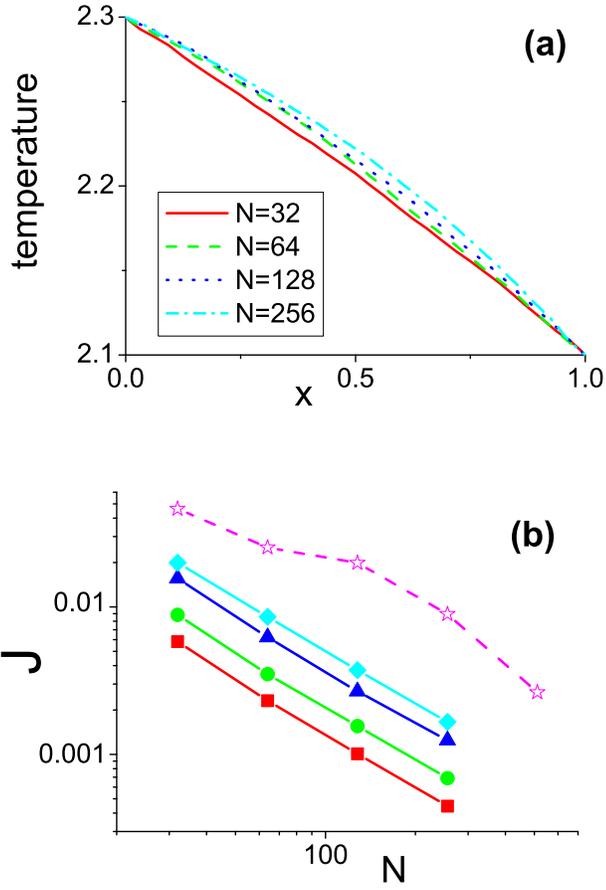}}
\caption{ The results of SRSs with random phase and periodic systems: 
{\bf (a)} Temperature profile of four chains, where $c=0.8$. 
{\bf (b)} Heat current $J$ versus $N$, from top to bottom, $c$ are $0.0$
 (periodic system), $0.2$, $0.4$, $0.6$ and $0.8$, respectively.
\label{fig3}}
\end{figure}

\begin{figure}
\centerline{\epsfxsize=10cm \epsfbox{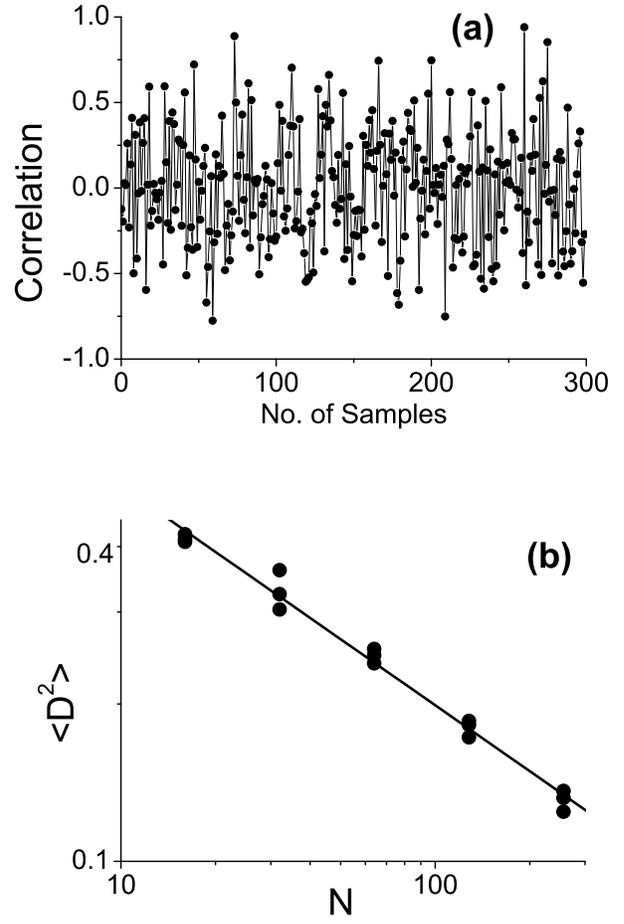}}
\caption{ The correlation of SRSs with random phase: {\bf (a)} The correlation
distribution of different samples, where $N=32$ and $c=0.2$. {\bf (b)} The 
fluctuation of correlation $\langle D^{2} \rangle$ versus $N$. We show
the results of three SRSs with different $c$ ($c=0.2, 0.5$ and $0.8$), 
they are almost indistinguishable. The solid line is the best fit one.
\label{fig4}}
\end{figure}

We also numerically simulate the heat conduction of SRS. Since the position 
disordered system is similar to the phase disordered system, 
we only show the results of the latter. The temperature profile ${\cal T}(x)$
and the $N$ dependence of $J_{u}$ in SRS are similar as that
in DRS, but ${\gamma} \approx 1.19$ in SRS, slightly larger than 
$1$ in DRS (seeing Fig. \ref{fig3}) and there is a slight difference on 
${\cal T}(x/N)$ of systems with different $N$. The differences may be due to 
the limiting chain length $N$ in our simulation. For different disordered 
magnitude $c$, the value of $\gamma$ is found to be indistinguishable. 
For comparison, we simulate the heat current $J_{u}$ in a periodic 
scattering system, the result is also not very clear. It may be necessary to 
simulate them in longer chains. In Fig. \ref{fig4}, we show the distribution 
of $D$ for different disordered realizations in SRS. The average of $D$ 
is about zero, in agreement with our expectation which the scattering 
disorders induce the absence of the correlation. Other interesting result is 
shown in Fig. \ref{fig4}(b): the fluctuation of $D$ decreases as $N$ 
increases, $\langle D^{2} \rangle = N^{-0.42}$. It indicates that 
$D \rightarrow 0$ in 
any single disordered realization of SRS while $N \rightarrow \infty$.
Contrarily, in periodic system, the correlation $D$ is not zero. 

\section{Discussions and Conclusion}

We have defined a spatical correlation along the 1D chains to describe 
scattering (i. e. transport) characteristic in the systems and 
theoretically expected that randomization of scattering determineed the 
existence of the long range spatial correlation, therefore properties of 
heat conduction. For DRS, the expectation was confirmed by both 
S-matrix theory and molecular dynamic simulation. 
For SRS and periodic system, according to the obtained correlations, normal 
and abnormal heat conductions were expected, respectively. 
But our numerical simulations in some small systems could not completely clear 
the expectation, more simulations might be necessary. 
A noted fact is that our model is very 
similar to the Ehrenfest gas channel~\cite{Li2002}. In the latter, the 
channel is quisal-1D systems with a small transverse coordinate (the 
height of the channel). Actually, the height corresponds to the `phase' 
in our models, a similar scattering theory can be derived. 
For the Ehrenfest channel, the results of Li {\it et al.}~\cite{Li2002}, 
which heat conduction is normal if the heights or positions of the 
scatterers are random and it is anormalous if their heights or positions are
periodic, can be easily understood from our results. In these cases, 
the required incoherent characteristic of motion to get normal heat conduction
can be ensured by the random `phases' and positions of scatterers, not 
needing additive dynamic chaos (non-zero Lyapunov Exponent).
Therefore, the conditions with normal heat conduction in 1D systems
are whether or not there are sufficient random to destory completely the 
correlation of motions along chains. The random may come from dynamic 
instablity, random scatterers or another sources, but no any particular 
random, such as chaos, is necessary. The judgement of the incoherent motions 
(i. e. normal transport) is whether or not the long range spatial 
correlations exist. 

 We also can consider another models, for example, an variant of 
ding-a-dong model, which its even-numbered particles are oscillatedly coupled
each other, rather than coupled with its individual lattice site in the 
initial model~\cite{Casati1984}. Due to the existence of some long-wave modes 
in the model, the motion correlation is obviously present, we expect the 
thermal conductivity should be abnormal. On general 1D systems, important 
questions are how to define suitable correlations and judge directly the 
existence of the correlations from the characteristic of systems.
For example, for momentum conservation systems, which 
thermal conductivities are anomalous if presure is not zero~\cite{Prosen2000}, 
it is worthy for connecting the conditions with existence of spatial 
correlations. 


X. Z. was supported by the Grants-in-Aid for Scientific Research of JSPS, his
current address is zhou@pe.titech.ac.jp. X. Z. thanks the help of Mr. C.-Q. 
Li.

\end{document}